\pdfoutput=1

\RequirePackage{fix-cm}

\documentclass[%
floatfix,
showkeys,
nofootinbib, %
superscriptaddress, %
]{revtex4-1}

\usepackage{cmap}

\usepackage{ucs}
\usepackage[utf8x]{inputenc}
\usepackage[T1,T2A]{fontenc}
\usepackage[german,russian,english]{babel}

\usepackage[sort&compress]{natbib}
\usepackage{amsmath}
\usepackage{amssymb}

\usepackage{mathtools}
\mathtoolsset{
showonlyrefs,
mathic = true
}

\allowdisplaybreaks

\usepackage{hyperref}
\hypersetup{backref,
 colorlinks=false}
\hypersetup{pdfborder=0 0 0}

\usepackage{microtype}
\UseMicrotypeSet[protrusion]{alltext}

\usepackage{graphicx}

\usepackage[scanall]{psfrag}

\usepackage{listings}
\usepackage{listingsutf8}
\usepackage{fixltx2e}

\usepackage{nicefrac}

\makeatletter
\def\ps@pprintTitle{%
     \let\@oddhead\@empty
     \let\@evenhead\@empty
     \let\@oddfoot\@empty
     \let\@evenfoot\@oddfoot}
\makeatother

\usepackage{physics}

\newcommand{\crd}[1]{\underline{\vphantom{j}{#1}}\,}

\begin{document}

\title{Algorithm for Lens Calculations in the Geometrized Maxwell Theory}

\author{M. N. Gevorkyan}
\email{gevorkyan_mn@rudn.university}
\affiliation{Department of Applied Probability and Informatics,\\
  Peoples' Friendship University of Russia (RUDN University),\\
  6 Miklukho-Maklaya str., Moscow, 117198, Russia}

\author{A. V. Demidova}
\email{demidova_av@rudn.university}
\affiliation{Department of Applied Probability and Informatics,\\
  Peoples' Friendship University of Russia (RUDN University),\\
  6 Miklukho-Maklaya str., Moscow, 117198, Russia}

\author{A. V. Korolkova}
\email{korolkova_av@rudn.university}
\affiliation{Department of Applied Probability and Informatics,\\
  Peoples' Friendship University of Russia (RUDN University),\\
  6 Miklukho-Maklaya str., Moscow, 117198, Russia}

\author{D. S. Kulyabov}
\email{kulyabov_ds@rudn.university}
\affiliation{Department of Applied Probability and Informatics,\\
  Peoples' Friendship University of Russia (RUDN University),\\
  6 Miklukho-Maklaya str., Moscow, 117198, Russia}
\affiliation{Laboratory of Information Technologies\\
  Joint Institute for Nuclear Research\\
  6 Joliot-Curie, Dubna, Moscow region, 141980, Russia}

\author{L. A. Sevastianov}
\email{sevastianov_la@rudn.university}
\affiliation{Department of Applied Probability and Informatics,\\
  Peoples' Friendship University of Russia (RUDN University),\\
  6 Miklukho-Maklaya str., Moscow, 117198, Russia}
\affiliation{Bogoliubov Laboratory of Theoretical Physics\\
  Joint Institute for Nuclear Research\\
  6 Joliot-Curie, Dubna, Moscow region, 141980, Russia}

\begin{abstract}
  Nowadays the geometric approach in optics is often used to find out
  media parameters based on propagation paths of the rays because in
  this case it is a direct problem. However inverse problem in the
  framework of geometrical optics is usually not given attention.

  The aim of this work is to demonstrate the work of the proposed the
  algorithm in the framework of geometrical approach to optics for
  solving the problem of finding the propagation path of the
  electromagnetic radiation depending on environmental parameters.
  The methods of differential geometry are used for effective metrics
  construction for isotropic and anisotropic media.  For effective
  metric space ray trajectories are obtained in the form of geodesic
  curves. The introduced algorithm is applied to well-known
  objects~--- Maxwell and Luneburg lenses. The similarity of results
  obtained by classical and geometric approach is demonstrated.
\end{abstract}


\maketitle
\section{Introduction}
\label{sec:intro}

Geometrical approach to Maxwell's equations has passed through several stages in
its development. Initial interest was caused by the General theory
of relativity. The works of L. I. Mandelstam,
I.~E.~Tamm~\cite{tamm:1924:jrpc::en,tamm:1925:jrpc::en,tamm:1925:mathann},
W.~Gordon~\cite{gordon:1923} belong to this period. In the absence of practical applications
the interest for this subject has gone. A new surge of interest arose during
the Golden age of the theory of relativity (1960--1975). 
The works of J.~Plebanski~\cite{plebanski:1960:electromagnetic_waves},
F.~Felice~\cite{felice:1971:as_optical_medium} belong to this period. However, it should be
note that in this period scientists failed to determine the application
of developed theory.

A new outbreak of interest in the geometric approach emerged in the mid
2000 as a side effect of the interest in
metamaterials~\cite{smolyaninov:2011:metamaterial}. We shell mention the studies of
J.~B.~Pendry~\cite{pendry:2006:controlling-em,schurig:2006:ray-tracing}
and U.~Leonhardt~\cite{leonhardt:2006:mapping,leonhardt:2009:light}.
These works gave rise to the whole direction~--- transformational
optics~\cite{foster:2015:spatial_transformations}.

For the classical approach to optics the direct and inverse problems
are usually formulated as follows:
\begin{itemize}
  \item the direct problem: from the environment settings to obtain
    the path of electromagnetic waves propagation;
  \item the inverse problem: from given propagations paths of
    electromagnetic waves to obtain environments parameters.
\end{itemize}

For the geometric approach to optics these problems are swapped:
\begin{itemize}
  \item the direct problem: for a given distribution paths of
    electromagnetic waves
    to obtain environment (medium) parameters;
  \item the inverse problem: from parameters of the environment to
    obtain the propagation path of electromagnetic waves.
\end{itemize}

Therefore, transformation optics works with direct problem of geometric optics (the inverse problem of classical optics).

Usually, in the framework of geometrical optics only  a direct problem is
solved, i.e. the task of finding the parameters of the environment from
the propagation path of electromagnetic waves. This framework does not
focuses on the inverse problem, i.e. finding the propagation path of
electromagnetic waves according to the known parameters of the medium.

This work aims to consistently present the algorithm of lenses
calculations with use of  geometrical optics approach and demonstrate
the convergence of our results with the results of the classical
optics approach.

The structure of this paper is following. In
section~\ref{sec:notation} the basic notation and
conventions used in the article are given. In section~\ref{sec:tamm:revert} we
describe the algorithm for inverse problem of geometrical optics.
In paragraph~\ref{sec:lenses} we present examples of calculation of
specific lenses. The results of numerical experiment are presented in
graphical form.

\section{Notations and conventions}
\label{sec:notation}

  \begin{enumerate}

  \item We will use the notation of abstract
    indices~\cite{penrose-rindler:spinors::en}. In this notation tensor
    as a
    complete object is denoted merely by an index (e.g., $x^{i}$). Its
    components are
    designated by underlined indices (e.g., $x^{\crd{i}}$).
    
  \item We will adhere to the following agreements. Greek indices
    ($\alpha$, $\beta$) will refer to the four-dimensional space, in the
    component form it looks like: $\crd{\alpha} = \overline{0,3}$.  Latin
    indices from the middle of the alphabet ($i$, $j$, $k$) will refer
    to the three-dimensional space, in the component form it looks like:
    $\crd {i} = \overline{1,3}$.
    


  \end{enumerate}

  \section{The algorithm of solving the inverse problem of geometrical optics}
  \label{sec:tamm:revert}

Although geometrical optics deals with the direct problem of
obtaining environmental parameters from rays propagation
trajectories, it is possible to solve the inverse problem:
calculation of lenses parameters.
 
Let us consider several options for solving the inverse problem
geometrical optics, namely, the cases of isotropic and anisotropic media.

We will use the following algorithm for solving the inverse problem.
\begin{enumerate}
\item Inputs are the environmental parameters such as the permittivity
  $\varepsilon_{i j}$ and the permeability $\mu_{i j}$ and the
  refractive index $n_{i j}$. If we take into account specifics of the
  geometrization based on quadratic metric, we have to consider only
  the refractive index.
\item From physical considerations, we choose ansatz for the effective
  metric tensor $g_{\alpha \beta}$.
\item Based on this ansatz we obtain the general form of effective
  metric tensor $g_{\alpha \beta}$. This process is iterative. There is no
  guarantee that the chosen ansatz will give the opportunity to obtain
  metric tensor. In this case we have to choose another ansatz.
\item  By substituting specific values of the parameters of the
  environment, we will receive a specific implementation of an
  effective metric tensor $g_{\alpha \beta}$.
\end{enumerate}

  Having an effective metric tensor, we can solve geometrical Maxwell
  equations and obtain the desired propagation path of
  electromagnetic
  waves~\cite{kulyabov:2016:pcs,kulyabov:2013:springer:cadabra}. In
  this paper we will use the geometric optics
  approximation~\cite{born-wolf:principles_optics::en,bruns:1895}. For
  this case, the rays will be propagated along the geodesic
  curve~\cite{ll:2::en,mtw::en}:
    \begin{equation}
  \dv[2]{x^{\gamma}}{t} +\Gamma_{\alpha \beta}^{\gamma}
  \dv{x^{\alpha}}{t} \dv{x^{\beta}}{t} = 0,
  \end{equation}
   where $x^{\gamma}(t)$ are coordinates of geodesic curve. The
   Christoffel symbols are defined as follows:

\begin{equation}
  \Gamma_{\alpha \beta}^{\gamma} =
  \frac{1}{2}
  g^{\gamma \delta}
  \qty(\pdv{g_{\alpha \delta}}{x^{\beta}} +
  \pdv{g_{\beta \delta}}{x^{\alpha}} -
  \pdv{g_{\alpha \beta}}{x^{\delta}}).
\end{equation}

  For calculations we use the following expressions for
  geometrized material equations:
\begin{equation}
  \label{eq:geom-maxwell:tamm:decart}
  \begin{gathered}
    D^{i} = \varepsilon^{i j} E_{j} + {}^{(1)}\gamma^{i}_{j} B^{j},
    \\
    H_{i} = (\mu^{-1})_{i j} B^{j} + {}^{(2)}\gamma^{j}_{i} E_{j}, \\
    \varepsilon^{\crd{i} \crd{j}} =
    -
    \sqrt{-g}
    \qty(g^{00} g^{\crd{i}\crd{j}} – g^{0\crd{i}} g^{0\crd{j}} )
    ,
    \\
    (\mu^{-1})_{\crd{i} \crd{j}} =
    \sqrt{-g}
    \varepsilon_{\crd{m}\crd{n}\crd{i}} \varepsilon_{\crd{k}\crd{l}\crd{j}} g^{\crd{n}\crd{k}} g^{\crd{m}\crd{l}},
    \\
    {}^{(1)}\gamma^{i}_{j} = {}^{(2)}\gamma^{i}_{j}
    =
    \sqrt{-g}
    \varepsilon_{\crd{k}\crd{l}\crd{j}} g^{0\crd{k}} g^{\crd{i}\crd{l}}.
  \end{gathered}
\end{equation}

Next, let us consider the isotropic and anisotropic cases.

  \subsection{Isotropic case}
\label{sec:tamm:revert:homogeneous}

Consider the implementation of the inverse problem of geometrization
for isotropic case. We will consider the case of diagonal metrics. All
spatial diagonal components of the metric tensor are equal to each
other. Thus, consider the following ansatz for the metric tensor:

\begin{equation}
  \label{eq:tamm:revert:homogeneous:g-ansatz}
  \begin{gathered}
    g_{\crd{\alpha} \crd{\beta}} =
    \mqty(\dmat[0]{a^2, - b^2, - b^2, - b^2}),
    \\
    g^{\crd{\alpha} \crd{\beta}} =
    \mqty(\dmat[0]{a^{-2}, - b^{-2}, - b^{-2}, - b^{-2}}),
    \\
    \sqrt{-g} = a b^{3}.
  \end{gathered}
\end{equation}

From relations~\eqref{eq:geom-maxwell:tamm:decart} we can write down the
expressions for the permittivity and the permeability:

\begin{equation}
  \label{eq:tamm:revert:homogeneous:epsilon-mu}
  \begin{gathered}
    \varepsilon^{\crd{i} \crd{j}}
    =
    a b^{3} a^{-2}
    \mqty(\dmat[0]{b^{-2}, b^{-2}, b^{-2}})
    =
    \frac{b}{a}
    \mqty(\dmat[0]{1,1,1})
    =
    \frac{b}{a}
    \delta^{\crd{i} \crd{j}}
    ,
    \\
    \qty( \mu^{-1} )_{\crd{i} \crd{j}}
    =
    a b^{3}
    \mqty(\dmat[0]{b^{-4}, b^{-4}, b^{-4}})
    =
    \frac{a}{b}
    \mqty(\dmat[0]{1,1,1})
    =
    \frac{a}{b}
    \delta_{\crd{i} \crd{j}}
    .
  \end{gathered}
\end{equation}

From equations~\eqref{eq:tamm:revert:homogeneous:epsilon-mu} we can write the
permittivity and the permeability in the following
form:
\begin{equation}
  \label{eq:tamm:revert:homogeneous:epsilon-mu:generic}
  \begin{gathered}
    \varepsilon^{\crd{i} \crd{j}} =
    \varepsilon  \delta^{\crd{i} \crd{j}},
    \quad
    \varepsilon =  \frac{b}{a},
    \\
    \qty( \mu^{-1} )_{\crd{i} \crd{j}} =
    \frac{1}{\mu}     \delta_{\crd{i} \crd{j}},
    \quad
    \mu = \frac{b}{a}.
  \end{gathered}
\end{equation}

  From~\eqref{eq:tamm:revert:homogeneous:epsilon-mu:generic} it is clear that
  remains one free parameter. Let the free parameter be
  $b$. Then let:
\begin{equation}
a = \frac{b}{\varepsilon}.
\end{equation}

Then, based on the ansatz~\eqref{eq:tamm:revert:homogeneous:g-ansatz}
we may write the metric tensor:
\begin{equation}
  \label{eq:tamm:revert:homogeneous:g:generic}
  g_{\crd{\alpha} \crd{\beta}} =
  \mqty(\dmat[0]{\frac{b^2}{\varepsilon^2}, - b^2, - b^2, - b^2}).
\end{equation}

  We will focus on Tamm~\cite{tamm:1924:jrpc::en,tamm:1925:jrpc::en,tamm:1925:mathann} approach. Let:
\begin{equation}
b^2 = \sqrt{\mu}.
\end{equation}
  Then the metric tensor~\eqref{eq:tamm:revert:homogeneous:g:generic}
  we may rewrite:
\begin{equation}
  \label{eq:tamm:revert:homogeneous:g:tamm:1}
  g_{\crd{\alpha} \crd{\beta}} =
  \mqty(\dmat[0]{\frac{\sqrt{\mu}}{\varepsilon^2},
    - \sqrt{\mu}, - \sqrt{\mu}, - \sqrt{\mu}}).
\end{equation}
  Or, considering the ratio:
\begin{equation}
  \varepsilon^{i j} \qty(\mu^{-1})_{j k} = \delta^{i}_{k},
\end{equation}
  it is possible to rewrite~\eqref{eq:tamm:revert:homogeneous:g:tamm:1} as
\begin{equation}
  \label{eq:tamm:revert:homogeneous:g:tamm:2}
  g_{\crd{\alpha} \crd{\beta}} =
  \mqty(\dmat[0]{\frac{1}{\varepsilon\sqrt{\mu}},
    - \sqrt{\mu}, - \sqrt{\mu}, - \sqrt{\mu}}).
\end{equation}

This ratio coincides with the solution proposed by
Tamm~\cite{tamm:1925:mathann}. The
expression~\eqref{eq:tamm:revert:homogeneous:g:tamm:1}
or~\eqref{eq:tamm:revert:homogeneous:g:tamm:2} sets the effective
geometry of the environment.

  \subsection{Anisotropic case}
  \label{sec:tamm:revert:nonhomogeneous}

Consider the simplest version of anisotropic medium. For this let us
consider the following ansatz for the metric tensor:
\begin{equation}
  \label{eq:tamm:revert:nonhomogeneous:g-ansatz}
  \begin{gathered}
    g_{\crd{\alpha} \crd{\beta}} =
    \mqty(\dmat[0]{(a_0)^2, - (a_1)^2, - (a_2)^2, - (a_3)^2}),
    \\
    g^{\crd{\alpha} \crd{\beta}} =
    \mqty(\dmat[0]{(a_0)^{-2}, - (a_1)^{-2}, - (a_2)^{-2}, - (a_3)^{-2}}),
    \\
    \sqrt{-g} = a_0 a_1 a_2 a_3.
  \end{gathered}
\end{equation}

From relations~\eqref{eq:geom-maxwell:tamm:decart} we may write down the expressions for
the permittivity and the permeability:
\begin{gather}
  \label{eq:tamm:revert:nonhomogeneous:epsilon-mu:epsilon}
    \varepsilon^{\crd{i} \crd{j}}
    =
    a_0 a_1 a_2 a_3
    (a_0)^{-2}
    \mqty(\dmat[0]{(a_1)^{-2}, (a_2)^{-2}, (a_3)^{-2}})
    =
    \mqty(\dmat[0]{\varepsilon_1, \varepsilon_1, \varepsilon_1})
    ,
    \\
    \label{eq:tamm:revert:nonhomogeneous:epsilon-mu:mu}
  \begin{multlined}
    \qty( \mu^{-1} )_{\crd{i} \crd{j}}
    = {} \\ {} =
    a_0 a_1 a_2 a_3
    \mqty(\dmat[0]{
      (a_2)^{-2} (a_3)^{-2},
      (a_3)^{-2} (a_1)^{-2},
      (a_1)^{-2} (a_2)^{-2}
    })
    = {} \\ {} =
    \mqty(\dmat[0]{\frac{1}{\mu_1},\frac{1}{\mu_2},\frac{1}{\mu_3}})
    .
  \end{multlined}
\end{gather}

From~\eqref{eq:tamm:revert:nonhomogeneous:epsilon-mu:epsilon}
and~\eqref{eq:tamm:revert:nonhomogeneous:epsilon-mu:mu} 
the permittivity and the permeability are obtained
as follows:
\begin{equation}
  \label{eq:tamm:revert:nonhomogeneous:epsilon-mu:2}
  \begin{gathered}
    \varepsilon_1 =
    \frac{\sqrt{-g}}{(a_{0})^2 (a_{1})^2},
    \quad
    \varepsilon_2 =
    \frac{\sqrt{-g}}{(a_{0})^2 (a_{2})^2},
    \quad
    \varepsilon_3 =
    \frac{\sqrt{-g}}{(a_{0})^2 (a_{3})^2},
    \\
    \mu_1 = \frac{(a_{2})^2 (a_{3})^2}{\sqrt{-g}},
    \quad
    \mu_2 = \frac{(a_{3})^2 (a_{1})^2}{\sqrt{-g}},
    \quad
    \mu_3 = \frac{(a_{1})^2 (a_{2})^2}{\sqrt{-g}}
    .
  \end{gathered}
\end{equation}

From~\eqref{eq:tamm:revert:nonhomogeneous:epsilon-mu:2} we may write
the following relations:
\begin{equation}
  \label{eq:tamm:revert:nonhomogeneous:epsilon-mu:3}
  \begin{gathered}
    \mu_2 \mu_3 = \frac{(a_{3})^2 (a_{1})^2}{\sqrt{-g}}
    \frac{(a_{1})^2 (a_{2})^2}{\sqrt{-g}}
    =
    \frac{(a_{1})^4}{\sqrt{-g}}
    \frac{(a_{2})^2 (a_{3})^2}{\sqrt{-g}}
    =
    \frac{(a_{1})^4}{\sqrt{-g}} \mu_1
    ,
    \\
    \mu_3 \mu_1 = \frac{(a_{1})^2 (a_{2})^2}{\sqrt{-g}}
    \frac{(a_{2})^2 (a_{3})^2}{\sqrt{-g}}
    =
    \frac{(a_{2})^4}{\sqrt{-g}}
    \frac{(a_{3})^2 (a_{1})^2}{\sqrt{-g}}
    =
    \frac{(a_{2})^4}{\sqrt{-g}} \mu_2
    ,
    \\
    \mu_1 \mu_2 = \frac{(a_{2})^2 (a_{3})^2}{\sqrt{-g}}
    \frac{(a_{3})^2 (a_{1})^2}{\sqrt{-g}}
    =
    \frac{(a_{3})^4}{\sqrt{-g}}
    \frac{(a_{1})^2 (a_{2})^2}{\sqrt{-g}}
    =
    \frac{(a_{3})^4}{\sqrt{-g}} \mu_3
    .
  \end{gathered}
\end{equation}

Let us write out the coefficients:
\begin{equation}
  \label{eq:tamm:revert:nonhomogeneous:epsilon-mu:4}
  \begin{gathered}
    a_1 = \sqrt{\frac{\mu_2 \mu_3}{\mu_1}},
    \\
    a_2 = \sqrt{\frac{\mu_3 \mu_1}{\mu_2}},
    \\
    a_3 = \sqrt{\frac{\mu_1 \mu_2}{\mu_3}}.
  \end{gathered}
\end{equation}

  Thus,~\eqref{eq:tamm:revert:nonhomogeneous:g-ansatz} is changed to:
\begin{equation}
  \label{eq:tamm:revert:nonhomogeneous:g:tamm:2}
  g_{\crd{\alpha} \crd{\beta}} =
  \mqty(\dmat[0]{
    \frac{1}{\sqrt{\varepsilon_1 \varepsilon_2 \mu_3}},
    - \sqrt{\frac{\mu_2 \mu_3}{\mu_1}},
    - \sqrt{\frac{\mu_3 \mu_1}{\mu_2}},
    - \sqrt{\frac{\mu_1 \mu_2}{\mu_3}}
  }).
\end{equation}

It is easy to see that in the isotropic case, the
ratio~\eqref{eq:tamm:revert:nonhomogeneous:g:tamm:2} proceeds in the
ratio of~\eqref{eq:tamm:revert:homogeneous:g:tamm:2}. The
expression~\eqref{eq:tamm:revert:nonhomogeneous:g:tamm:2} sets the
effective geometry of the environment.

  \section{Examples of lenses calculation in geometrical optics}
  \label{sec:lenses}
  For examples of calculations we use a widely known
  Maxwell (fish-eye)~\cite{maxwell:1854:fish-eye}, and
  Luneburg~\cite{luneburg:1964} lenses.
  Also, these lenses are important because for them one may obtain
  analytical solutions. In addition, convergence of solutions in
  the classical and geometrical approaches could be used for verification.

\subsection{Maxwell lens}

Maxwell lens~\cite{maxwell:1854:fish-eye} is constructed so that, in
special case, when the rays emit from a point source located one
side of the lens, they are focused at one point on the opposite
side of the lens.

The refractive index $n$ changes from
$2 n_{0}$ in the centre up to $n_{0}$ at the surface:
\begin{equation}
  n(r) =
  \begin{dcases}
    \frac{2 n_{0}}{ 1 +  \qty(\frac{r}{R})^2}, & r \leqslant R,\\
    n_{0}, & r > R.
  \end{dcases}
\end{equation}
Here $R$ is the radius of the sphere or cylinder. Also
usually one consider $n_0 = 1$.
Since in the method of the geometrization based on quadratic metric
the permittivity and the permeability are equal, so we can write:
\begin{equation}
  \begin{gathered}
    \varepsilon^{\crd{i} \crd{j}} = \mu^{\crd{i} \crd{j}},
    \\
    \varepsilon^{\crd{i} \crd{j}} = \varepsilon \delta^{\crd{i} \crd{j}},
    \quad \mu^{\crd{i} \crd{j}} = \mu \delta^{\crd{i} \crd{j}},
    \\
    n = \sqrt{\varepsilon \mu} = \varepsilon = \mu,
    \\
    \varepsilon = \mu = \frac{2}{ 1 +  \qty(\frac{r}{R})^2}, \quad
    r \leqslant R, n_0 := 1.
  \end{gathered}
\end{equation}

Then, from~\eqref{eq:tamm:revert:homogeneous:g:tamm:2}, we get the following metric:
\begin{equation}
  \label{fig:design:luneburg:g}
  g_{\crd{\alpha} \crd{\beta}} =
  \mqty(\dmat[0]{
    \qty(\frac{2}{ 1 +  \qty(\frac{r}{R})^2})^{-3/2},
    - \qty(\frac{2}{ 1 +  \qty(\frac{r}{R})^2})^{1/2},
    - \qty(\frac{2}{ 1 +  \qty(\frac{r}{R})^2})^{1/2},
    - \qty(\frac{2}{ 1 +  \qty(\frac{r}{R})^2})^{1/2}
  }).
\end{equation}

\begin{figure}
  \centering
  \includegraphics[width=0.4\textwidth]{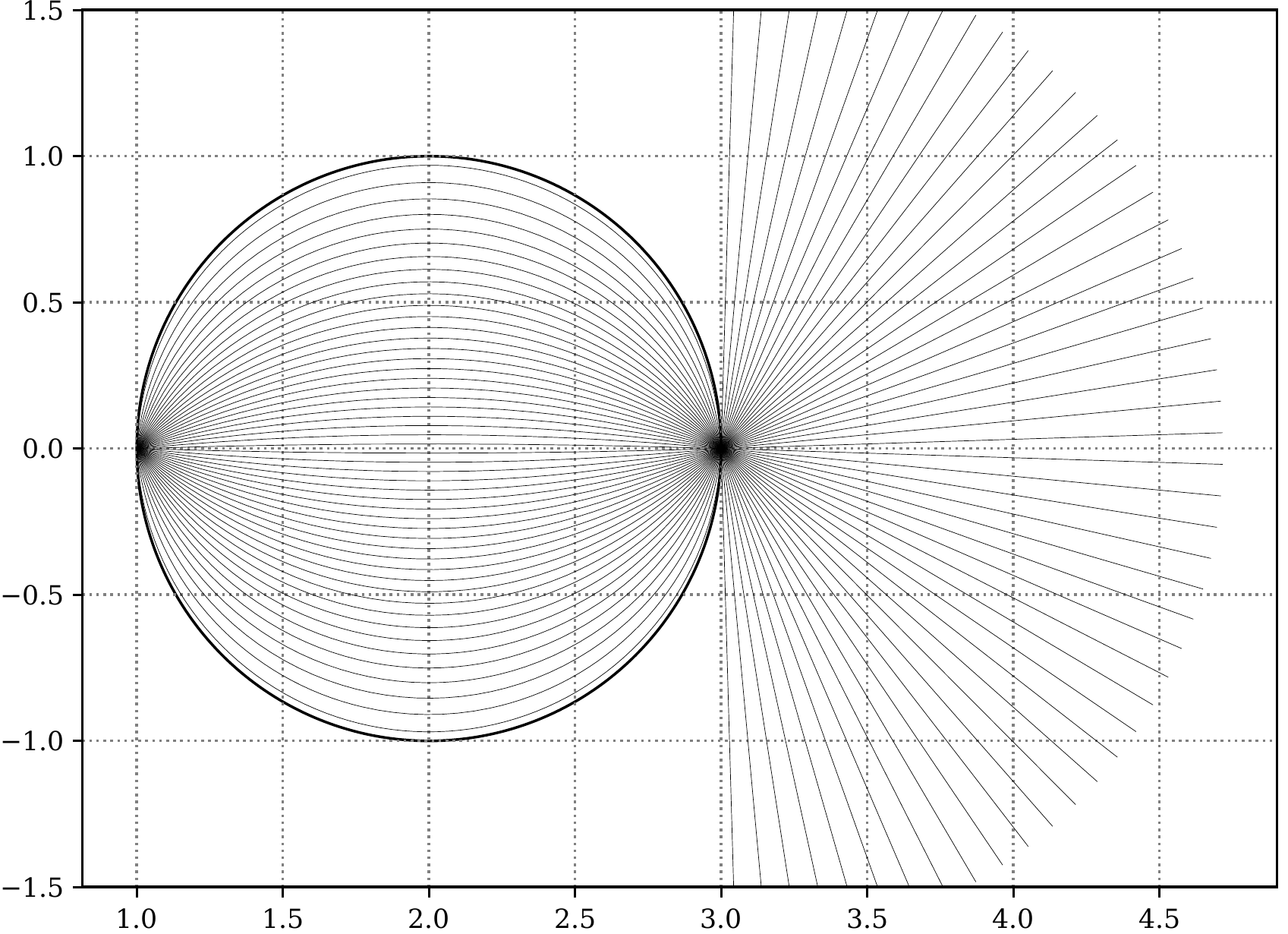}
  \caption{Ray trajectories as geodesic curves for Maxwell lens}
\label{fig:design:maxwell:geodesic:rays}
\end{figure}

Then we can depict the trajectories of rays as geodesic curves in this
space (see Fig.~\ref{fig:design:maxwell:geodesic:rays}). This figure
shows that the behavior of the trajectories of the rays coincides with
the theoretically predicted results from classical
optics~\cite{maxwell:1854:fish-eye}, that is, the rays, emerging from
source on the surface of the lens, are focused at a point located on the the
opposite surface of the lens.

  \subsection{Luneburg Lens}

Luneburg lens~\cite{luneburg:1964,morgan:1958:luneberg_lens} is
the gradient lens. The refractive index changes depending on
the distance from the center (spherical lens) or from the axis (cylindrical
lens).
With the passage of the lens the parallel rays are focused at one point on
the surface of the lens. The rays emitted by a point source on the surface
lenses form a parallel beam.

The refractive index $n$ changes from
$\sqrt{2} n_{0}$ in the centre up to $n_{0}$ at the surface:
\begin{equation}
    n(r) =
    \begin{dcases}
      n_{0} \sqrt{2-\qty(\frac{r}{R})^2}, & r \leqslant R,\\
      n_{0}, & r > R.
    \end{dcases}
  \end{equation}
Here $R$ is the radius of the sphere or cylinder. Also,
usually one consider $n_0 = 1$.

Since in the method of the geometrization based on quadratic metric
the permittivity and the permeability are equal, we can write:
\begin{equation}
  \begin{gathered}
    \varepsilon^{\crd{i} \crd{j}} = \mu^{\crd{i} \crd{j}},
    \\
    \varepsilon^{\crd{i} \crd{j}} = \varepsilon \delta^{\crd{i} \crd{j}},
    \quad \mu^{\crd{i} \crd{j}} = \mu \delta^{\crd{i} \crd{j}},
    \\
    n = \sqrt{\varepsilon \mu} = \varepsilon = \mu,
    \\
    \varepsilon = \mu = \sqrt{2-\qty(\frac{r}{R})^2}, \quad
    r \leqslant R, n_0 := 1.
  \end{gathered}
\end{equation}

Then, from~\eqref{eq:tamm:revert:homogeneous:g:tamm:2}, we get the following metric:

\begin{equation}
  \label{fig:design:luneburg:g}
  g_{\crd{\alpha} \crd{\beta}} =
  \mqty(\dmat[0]{\qty(2-\frac{r}{R})^{-3/4},
    - \qty(2-\frac{r}{R})^{1/4}, - \qty(2-\frac{r}{R})^{1/4}, - \qty(2-\frac{r}{R})^{1/4}}).
\end{equation}

\begin{figure}
  \centering
  \includegraphics[width=0.4\textwidth]{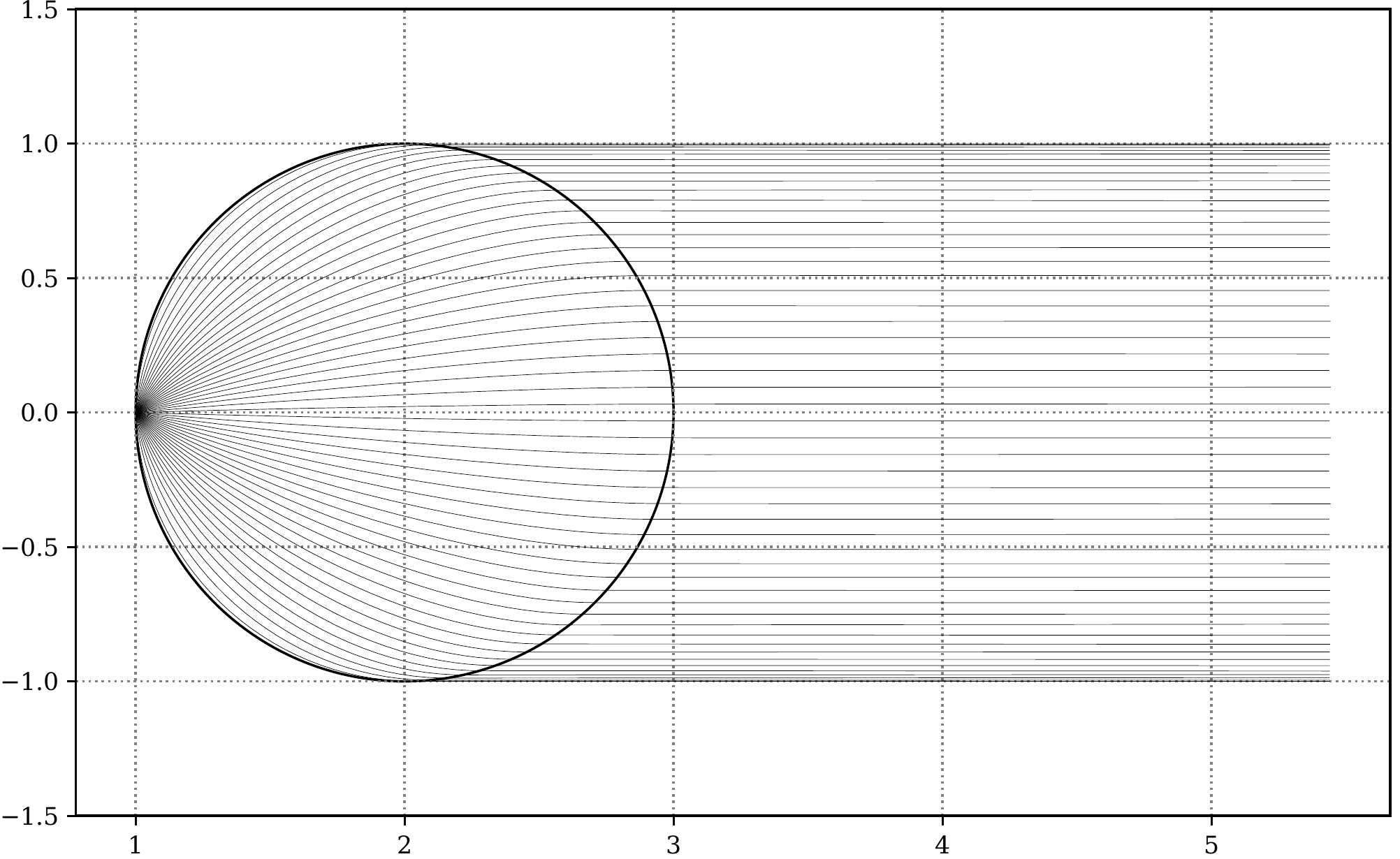}
  \caption{Ray trajectories as geodesic curves for Luneburg lens}
\label{fig:design:luneburg:geodesic:rays}
\end{figure}

  Then we can depict the trajectories of rays as geodesic curves in
  this space (see Fig.~\ref{fig:design:luneburg:geodesic:rays}). This
  figure shows that the behavior of the trajectories of the rays
  coincides with the the theoretically prediction of classical
  optics~\cite{morgan:1958:luneberg_lens}. The rays, which emerge from
  a point source on the surface of the lens, form a parallel beam.

\section{Conclusion}
\label{sec:conclusion}

The authors proposed the algorithm for the calculation of the lenses in the framework of
the geometrical approach to optics. The ansatzes for cases of
isotropic and anisotropic media are proposed. For example, the widely
known Luneburg and Maxwell lenses demonstrate a coincidence of the
classical and geometric approaches. For simplicity
the calculations were performed only for geometrical optics.

Unfortunately, it is not clear if the proposed approach for lens calculation
on the basis of geometrical optics has any advantages over classical one. This
question will be the subject of further research.

\begin{acknowledgments}

The work is partially supported by RFBR grants No's~15-07-08795 and 16-07-00556.
Also the publication was prepared with the support of the
``RUDN University Program 5-100''.

\end{acknowledgments}

 \bibliographystyle{elsarticle-num}

\bibliography{bib/lenses-calculation-algorithm/cite}

\end{document}


\title{Алгоритм расчёта линз в геометризованной теории Максвелла}

\author{М. Н. Геворкян}
\email{gevorkyan_mn@rudn.university}
\affiliation{Кафедра прикладной информатики и теории вероятностей,\\
  Российский университет дружбы народов,\\
  ул. Миклухо-Маклая, д. 6, Москва, Россия, 117198}

\author{А. В. Демидова}
\email{demidova_av@rudn.university}
\affiliation{Кафедра прикладной информатики и теории вероятностей,\\
  Российский университет дружбы народов,\\
  ул. Миклухо-Маклая, д. 6, Москва, Россия, 117198}

\author{А. В. Королькова}
\email{korolkova_av@rudn.university}
\affiliation{Кафедра прикладной информатики и теории вероятностей,\\
  Российский университет дружбы народов,\\
  ул. Миклухо-Маклая, д. 6, Москва, Россия, 117198}

\author{Д. С. Кулябов}
\email{kulyabov_ds@rudn.university}
\affiliation{Кафедра прикладной информатики и теории вероятностей,\\
  Российский университет дружбы народов,\\
  ул. Миклухо-Маклая, д. 6, Москва, Россия, 117198}
\affiliation{Лаборатория информационных технологий,\\
  Объединённый институт ядерных исследований,\\
  ул. Жолио-Кюри 6, Дубна, Московская область, Россия, 141980}

\author{Л. А. Севастьянов}
\email{sevastianov_la@rudn.university}
\affiliation{Кафедра прикладной информатики и теории вероятностей,\\
  Российский университет дружбы народов,\\
  ул. Миклухо-Маклая, д. 6, Москва, Россия, 117198}
\affiliation{Лаборатория теоретической физики,\\
  Объединённый институт ядерных исследований,\\
  ул. Жолио-Кюри 6, Дубна, Московская область, Россия, 141980}

\begin{abstract}
Для нахождения параметров среды по траекториям распространения
электромагнитного излучения всё чаще используется геометрический подход к
оптике, поскольку в данном подходе эта задача является прямой. Однако
обратной задаче в рамках геометризованной оптики обычно не уделяется
внимания.

Целью данной работы  является демонстрация предлагаемого в рамках
геометрического подхода к оптике
алгоритма решения задачи нахождения траектории распространения электромагнитного
излучения в зависимости от параметров среды. Для получения алгоритма авторами сформулирована
обратная задача геометрического подхода к оптике.
С помощью методов
дифференциальной геометрии построены метрики эффективного пространства
для изотропной и анизотропной среды. Траектории лучей получены в форме
геодезических (приближение геометрической оптики) в эффективном
пространстве. Алгоритм применён для нахождения траекторий лучей для
широко известных объектов~--- линз Максвелла и
Люнеберга. Продемонстрировано совпадение результатов геометризованной
и классической оптики.
\end{abstract}

  \keywords{уравнения Максвелла, риманова метрика, расслоенные
    пространства, линза Максвелла, линза Люнеберга}

\maketitle

\section{Введение}
\label{sec:intro}

Геометрический подход к уравнениям Максвелла прошёл несколько этапов в
своём развитии. Первоначальный интерес был вызван общей теорией
относительности. Это работы Л. И. Мандельштама,
И. Е. Тамма~\cite{tamm:1924:jrpc::ru,tamm:1925:jrpc::ru,tamm:1925:mathann},
В. Гордона~\cite{gordon:1923}. В отсутствии практических приложений
интерес к данной теме угас. Новая вспышка интереса возникла во время
золотого века теории относительности (1960--1975~гг.). Это работы
Е. Плебаньского~\cite{plebanski:1960:electromagnetic_waves}, Ф. де
Феличе~\cite{felice:1971:as_optical_medium}. Впрочем, следует
заметить, что и в этот период не удалось определиться с приложениями
данной теории.

Новая вспышка интереса к геометрическому подходу проявилась в середине
2000-х годов на фоне интереса к
метаматериалам~\cite{smolyaninov:2011:metamaterial}. Это работы
Д. Б. Пендри~\cite{pendry:2006:controlling-em,schurig:2006:ray-tracing}
и У. Леонгардта~\cite{leonhardt:2006:mapping,leonhardt:2009:light}.
Эти работы породили целое направление~--- трансформационную
оптику~\cite{foster:2015:spatial_transformations}.

В классическом подходе к оптике обычно прямая и обратная задача
формулируются следующим образом:
\begin{itemize}
\item прямая задача: по параметрам среды получить пути распространения
  электромагнитных волн;
\item обратная задача: по заданным путям распространения
  электромагнитных волн получить параметры среды.
\end{itemize}

При геометрическом подходе к оптике эти задачи меняются местами:
\begin{itemize}
\item прямая задача: по заданным путям распространения
  электромагнитных волн получить параметры среды;
\item обратная задача: по параметрам среды получить путь распространения
  электромагнитных волн.
\end{itemize}

Таким образом, трансформационная оптика работает с прямой задачей
геометризованной оптики (обратной задачей классической оптики).

Обычно при использовании геометризованной оптики решают только прямую
задачу, то есть задачу нахождения параметров среды по траекториям
распространения электромагнитных волн. При этом практически не
уделяется внимание обратной задаче, то есть нахождению траекторий
распространения электромагнитных волн по известным параметрам среды.

Работа ставит целью последовательно изложить алгоритм расчёта линз в
геометризованной оптике и продемонстрировать совпадение результатов
его применения с
результатами классического подхода к оптике.

Структура работы следующая. В~разделе~\ref{sec:notation} даются основные обозначения и соглашения,
применяемые в данной работе. В разделе~\ref{sec:tamm:revert}
описывается собственно алгоритм для обратной задачи геометризованной
оптики. В разделе~\ref{sec:lenses} приведены примеры расчёта
конкретных линз. Результаты численного эксперимента продемонстрированы
на иллюстрациях.

\section{Обозначения и соглашения}

\label{sec:notation}

  \begin{enumerate}

  \item Будем использовать нотацию абстрактных
    индексов~\cite{penrose-rindler:spinors::ru}. В данной нотации тензор как
    целостный объект обозначается просто индексом (например, $x^{i}$),
    компоненты обозначаются подчёркнутым индексом (например,
    $x^{\crd{i}}$).

  \item Будем придерживаться следующих соглашений.  Греческие индексы
    ($\alpha$, $\beta$) будут относиться к четырёхмерному
    пространству и в компонентном виде будут иметь следующие значения:
    $\crd{\alpha} = \overline{0,3}$. Латинские индексы из середины
    алфавита ($i$, $j$, $k$) будут относиться к трёхмерному
    пространству и в компонентном виде будут иметь следующие значения:
    $\crd{i} = \overline{1,3}$.



  \end{enumerate}

  \section{Алгоритм решения обратной задачи геометризованной оптики}
  \label{sec:tamm:revert}

  Хотя геометризованная оптика имеет в качестве прямой задачи получение
  параметров среды по траекториям распространения лучей, возможно и
  решение обратной задачи. То есть с помощью геометризованной оптики
  возможен и расчёт линз.

  Рассмотрим несколько вариантов решения обратной задачи
  геометризованной оптики, а именно случаи изотропной и анизотропной сред.

  Будем использовать следующий алгоритм решения обратной задачи.
  \begin{enumerate}
  \item Входными данными для нас являются параметры среды, такие как
    диэлектрическая проницаемость $\varepsilon_{i j}$, магнитная
    проницаемость $\mu_{i j}$, показатель преломления $n_{i j}$. В связи
    со спецификой геометризации на основе квадратичной метрики реально
    учитывается только показатель преломления.
  \item Исходя из физических соображений, подбирается анзац для эффективного
    метрического тензора $g_{\alpha \beta}$.
  \item На основании анзаца получаем общий вид эффективного
    метрического тензора $g_{\alpha \beta}$. Данный процесс итеративный. Нет никакой
    гарантии, что наперёд заданный анзац даст возможность получить
    метрический тензор. В этом случае следует заменить анзац.

  \item Подставляя конкретные значения параметров среды, получим
    конкретную реализацию эффективного метрического тензора $g_{\alpha \beta}$.
  \end{enumerate}

  Имея эффективный метрический тензор, можно решить геометризованные
  уравнения Максвелла и получить искомые траектории распространения
  электромагнитных
  волн~\cite{kulyabov:2016:pcs,kulyabov:2013:springer:cadabra}. В
  рамках статьи упростим задачу и будем использовать приближение
  геометрической
  оптики~\cite{born-wolf:principles_optics::ru,bruns:1895}. В этом
  случае лучи будут распространяться по
  геодезическим~\cite{ll:2::ru,mtw:1::ru}:
    \begin{equation}
    \dv[2]{x^{\gamma}}{t} +\Gamma_{\alpha \beta}^{\gamma}
    \dv{x^{\alpha}}{t} \dv{x^{\beta}}{t} = 0,
  \end{equation}
  где $x^{\gamma}(t)$~--- координаты геодезической. Символы
  Кристоффеля задаются следующим образом:

\begin{equation}
  \Gamma_{\alpha \beta}^{\gamma} =
  \frac{1}{2}
  g^{\gamma \delta}
  \qty(\pdv{g_{\alpha \delta}}{x^{\beta}} +
  \pdv{g_{\beta \delta}}{x^{\alpha}} -
  \pdv{g_{\alpha \beta}}{x^{\delta}}).
\end{equation}

  Для расчётов будем использовать следующие выражения для
  геометризованных материальных уравнений:
\begin{equation}
  \label{eq:geom-maxwell:tamm:decart}
  \begin{gathered}
    D^{i} = \varepsilon^{i j} E_{j} + {}^{(1)}\gamma^{i}_{j} B^{j},
    \\
    H_{i} = (\mu^{-1})_{i j} B^{j} + {}^{(2)}\gamma^{j}_{i} E_{j}, \\
    \varepsilon^{\crd{i} \crd{j}} =
    -
    \sqrt{-g}
    \qty(g^{00} g^{\crd{i}\crd{j}} – g^{0\crd{i}} g^{0\crd{j}} )
    ,
    \\
    (\mu^{-1})_{\crd{i} \crd{j}} =
    \sqrt{-g}
    \varepsilon_{\crd{m}\crd{n}\crd{i}} \varepsilon_{\crd{k}\crd{l}\crd{j}} g^{\crd{n}\crd{k}} g^{\crd{m}\crd{l}},
    \\
    {}^{(1)}\gamma^{i}_{j} = {}^{(2)}\gamma^{i}_{j}
    =
    \sqrt{-g}
    \varepsilon_{\crd{k}\crd{l}\crd{j}} g^{0\crd{k}} g^{\crd{i}\crd{l}}.
  \end{gathered}
\end{equation}

Далее перейдём к рассмотрению изотропного и анизотропного случаев.

  \subsection{Изотропный случай}
\label{sec:tamm:revert:homogeneous}

Рассмотрим реализацию обратной задачи геометризации для изотропного
случая. Будем рассматривать случай диагональной метрики. Все
пространственные диагональные компоненты метрического тензора равны
между собой. Таким образом, рассмотрим следующий анзац для
метрического тензора:

\begin{equation}
  \label{eq:tamm:revert:homogeneous:g-ansatz}
  \begin{gathered}
    g_{\crd{\alpha} \crd{\beta}} =
    \mqty(\dmat[0]{a^2, - b^2, - b^2, - b^2}),
    \\
    g^{\crd{\alpha} \crd{\beta}} =
    \mqty(\dmat[0]{a^{-2}, - b^{-2}, - b^{-2}, - b^{-2}}),
    \\
    \sqrt{-g} = a b^{3}.
  \end{gathered}
\end{equation}

Из соотношений~\eqref{eq:geom-maxwell:tamm:decart} запишем выражения для
диэлектрической и магнитной проницаемостей:

\begin{equation}
  \label{eq:tamm:revert:homogeneous:epsilon-mu}
  \begin{gathered}
    \varepsilon^{\crd{i} \crd{j}}
    =
    a b^{3} a^{-2}
    \mqty(\dmat[0]{b^{-2}, b^{-2}, b^{-2}})
    =
    \frac{b}{a}
    \mqty(\dmat[0]{1,1,1})
    =
    \frac{b}{a}
    \delta^{\crd{i} \crd{j}}
    ,
    \\
    \qty( \mu^{-1} )_{\crd{i} \crd{j}}
    =
    a b^{3}
    \mqty(\dmat[0]{b^{-4}, b^{-4}, b^{-4}})
    =
    \frac{a}{b}
    \mqty(\dmat[0]{1,1,1})
    =
    \frac{a}{b}
    \delta_{\crd{i} \crd{j}}
    .
  \end{gathered}
\end{equation}

Из~\eqref{eq:tamm:revert:homogeneous:epsilon-mu} запишем проницаемости
в следующем виде:
\begin{equation}
  \label{eq:tamm:revert:homogeneous:epsilon-mu:generic}
  \begin{gathered}
    \varepsilon^{\crd{i} \crd{j}} =
    \varepsilon  \delta^{\crd{i} \crd{j}},
    \quad
    \varepsilon =  \frac{b}{a},
    \\
    \qty( \mu^{-1} )_{\crd{i} \crd{j}} =
    \frac{1}{\mu}     \delta_{\crd{i} \crd{j}},
    \quad
    \mu = \frac{b}{a}.
  \end{gathered}
\end{equation}

Из~\eqref{eq:tamm:revert:homogeneous:epsilon-mu:generic} видно, что у
нас остаётся один свободный параметр. Пусть свободным параметром будет
$b$. Тогда положим:
\begin{equation}
a = \frac{b}{\varepsilon}.
\end{equation}

  Тогда на основе анзаца~\eqref{eq:tamm:revert:homogeneous:g-ansatz}
  запишем метрический тензор:
\begin{equation}
  \label{eq:tamm:revert:homogeneous:g:generic}
  g_{\crd{\alpha} \crd{\beta}} =
  \mqty(\dmat[0]{\frac{b^2}{\varepsilon^2}, - b^2, - b^2, - b^2}).
\end{equation}

  Будем ориентироваться на идеи
  Тамма~\cite{tamm:1924:jrpc::ru,tamm:1925:jrpc::ru,tamm:1925:mathann}.
  Положим:
\begin{equation}
b^2 = \sqrt{\mu}.
\end{equation}
  Тогда~\eqref{eq:tamm:revert:homogeneous:g:generic} запишется следующим образом:
\begin{equation}
  \label{eq:tamm:revert:homogeneous:g:tamm:1}
  g_{\crd{\alpha} \crd{\beta}} =
  \mqty(\dmat[0]{\frac{\sqrt{\mu}}{\varepsilon^2},
    - \sqrt{\mu}, - \sqrt{\mu}, - \sqrt{\mu}}).
\end{equation}
  Или, учитывая соотношение:
\begin{equation}
  \varepsilon^{i j} \qty(\mu^{-1})_{j k} = \delta^{i}_{k},
\end{equation}
можно переписать~\eqref{eq:tamm:revert:homogeneous:g:tamm:1} как
\begin{equation}
  \label{eq:tamm:revert:homogeneous:g:tamm:2}
  g_{\crd{\alpha} \crd{\beta}} =
  \mqty(\dmat[0]{\frac{1}{\varepsilon\sqrt{\mu}},
    - \sqrt{\mu}, - \sqrt{\mu}, - \sqrt{\mu}}).
\end{equation}

Это соотношение совпадает с решением, предложенным
Таммом~\cite{tamm:1925:mathann}. Выражение~\eqref{eq:tamm:revert:homogeneous:g:tamm:1}
или~\eqref{eq:tamm:revert:homogeneous:g:tamm:2} задаёт эффективную
геометрию исследуемого пространства.

  \subsection{Анизотропный случай}
  \label{sec:tamm:revert:nonhomogeneous}

Рассмотрим простейший вариант анизотропной среды.  Для этого
рассмотрим следующий анзац для метрического тензора:
\begin{equation}
  \label{eq:tamm:revert:nonhomogeneous:g-ansatz}
  \begin{gathered}
    g_{\crd{\alpha} \crd{\beta}} =
    \mqty(\dmat[0]{(a_0)^2, - (a_1)^2, - (a_2)^2, - (a_3)^2}),
    \\
    g^{\crd{\alpha} \crd{\beta}} =
    \mqty(\dmat[0]{(a_0)^{-2}, - (a_1)^{-2}, - (a_2)^{-2}, - (a_3)^{-2}}),
    \\
    \sqrt{-g} = a_0 a_1 a_2 a_3.
  \end{gathered}
\end{equation}

  Из соотношений~\eqref{eq:geom-maxwell:tamm:decart} запишем выражения для
  диэлектрической и магнитной проницаемостей:
\begin{gather}
  \label{eq:tamm:revert:nonhomogeneous:epsilon-mu:epsilon}
    \varepsilon^{\crd{i} \crd{j}}
    =
    a_0 a_1 a_2 a_3
    (a_0)^{-2}
    \mqty(\dmat[0]{(a_1)^{-2}, (a_2)^{-2}, (a_3)^{-2}})
    =
    \mqty(\dmat[0]{\varepsilon_1, \varepsilon_1, \varepsilon_1})
    ,
    \\
    \label{eq:tamm:revert:nonhomogeneous:epsilon-mu:mu}
  \begin{multlined}
    \qty( \mu^{-1} )_{\crd{i} \crd{j}}
    = {} \\ {} =
    a_0 a_1 a_2 a_3
    \mqty(\dmat[0]{
      (a_2)^{-2} (a_3)^{-2},
      (a_3)^{-2} (a_1)^{-2},
      (a_1)^{-2} (a_2)^{-2}
    })
    = {} \\ {} =
    \mqty(\dmat[0]{\frac{1}{\mu_1},\frac{1}{\mu_2},\frac{1}{\mu_3}})
    .
  \end{multlined}
\end{gather}

Из~\eqref{eq:tamm:revert:nonhomogeneous:epsilon-mu:epsilon}
и~\eqref{eq:tamm:revert:nonhomogeneous:epsilon-mu:mu} запишем
проницаемости в следующем виде:
\begin{equation}
  \label{eq:tamm:revert:nonhomogeneous:epsilon-mu:2}
  \begin{gathered}
    \varepsilon_1 =
    \frac{\sqrt{-g}}{(a_{0})^2 (a_{1})^2},
    \quad
    \varepsilon_2 =
    \frac{\sqrt{-g}}{(a_{0})^2 (a_{2})^2},
    \quad
    \varepsilon_3 =
    \frac{\sqrt{-g}}{(a_{0})^2 (a_{3})^2},
    \\
    \mu_1 = \frac{(a_{2})^2 (a_{3})^2}{\sqrt{-g}},
    \quad
    \mu_2 = \frac{(a_{3})^2 (a_{1})^2}{\sqrt{-g}},
    \quad
    \mu_3 = \frac{(a_{1})^2 (a_{2})^2}{\sqrt{-g}}
    .
  \end{gathered}
\end{equation}

Из~\eqref{eq:tamm:revert:nonhomogeneous:epsilon-mu:2} выпишем
следующие соотношения:
\begin{equation}
  \label{eq:tamm:revert:nonhomogeneous:epsilon-mu:3}
  \begin{gathered}
    \mu_2 \mu_3 = \frac{(a_{3})^2 (a_{1})^2}{\sqrt{-g}}
    \frac{(a_{1})^2 (a_{2})^2}{\sqrt{-g}}
    =
    \frac{(a_{1})^4}{\sqrt{-g}}
    \frac{(a_{2})^2 (a_{3})^2}{\sqrt{-g}}
    =
    \frac{(a_{1})^4}{\sqrt{-g}} \mu_1
    ,
    \\
    \mu_3 \mu_1 = \frac{(a_{1})^2 (a_{2})^2}{\sqrt{-g}}
    \frac{(a_{2})^2 (a_{3})^2}{\sqrt{-g}}
    =
    \frac{(a_{2})^4}{\sqrt{-g}}
    \frac{(a_{3})^2 (a_{1})^2}{\sqrt{-g}}
    =
    \frac{(a_{2})^4}{\sqrt{-g}} \mu_2
    ,
    \\
    \mu_1 \mu_2 = \frac{(a_{2})^2 (a_{3})^2}{\sqrt{-g}}
    \frac{(a_{3})^2 (a_{1})^2}{\sqrt{-g}}
    =
    \frac{(a_{3})^4}{\sqrt{-g}}
    \frac{(a_{1})^2 (a_{2})^2}{\sqrt{-g}}
    =
    \frac{(a_{3})^4}{\sqrt{-g}} \mu_3
    .
  \end{gathered}
\end{equation}

Выразим коэффициенты:
\begin{equation}
  \label{eq:tamm:revert:nonhomogeneous:epsilon-mu:4}
  \begin{gathered}
    a_1 = \sqrt{\frac{\mu_2 \mu_3}{\mu_1}},
    \\
    a_2 = \sqrt{\frac{\mu_3 \mu_1}{\mu_2}},
    \\
    a_3 = \sqrt{\frac{\mu_1 \mu_2}{\mu_3}}.
  \end{gathered}
\end{equation}

  Таким образом~\eqref{eq:tamm:revert:nonhomogeneous:g-ansatz} запишется следующим образом:
\begin{equation}
  \label{eq:tamm:revert:nonhomogeneous:g:tamm:2}
  g_{\crd{\alpha} \crd{\beta}} =
  \mqty(\dmat[0]{
    \frac{1}{\sqrt{\varepsilon_1 \varepsilon_2 \mu_3}},
    - \sqrt{\frac{\mu_2 \mu_3}{\mu_1}},
    - \sqrt{\frac{\mu_3 \mu_1}{\mu_2}},
    - \sqrt{\frac{\mu_1 \mu_2}{\mu_3}}
  }).
\end{equation}

Легко видеть, что в изотропном случае
соотношение~\eqref{eq:tamm:revert:nonhomogeneous:g:tamm:2} переходит в
соотношение~\eqref{eq:tamm:revert:homogeneous:g:tamm:2}. Выражение~\eqref{eq:tamm:revert:nonhomogeneous:g:tamm:2}
задаёт эффективную геометрию исследуемого пространства.

  \section{Примеры расчёта линз в геометризованной оптике}
  \label{sec:lenses}

  Для примеров расчётов нами выбраны широко известные объекты: линза
  Максвелла (<<рыбий глаз>>)~\cite{maxwell:1854:fish-eye} и линза
  Люнеберга~\cite{luneburg:1964}.
  Также эти объекты привлекательны тем, что позволяют получить
  аналитические выражения для решений. Кроме того, совпадение решений в
  случае классического и геометризованного подходов может служить целям верификации.

\subsection{Линза Максвелла}

Линза Максвелла~\cite{maxwell:1854:fish-eye} строится так, что, в
частном случае,
при испускании лучей из точечного источника, находящегося с одной
стороны линзы, лучи фокусируются в одной точке на противоположной
поверхности линзы.

Показатель преломления $n$ меняется от
$2 n_{0}$  в центре до $n_{0}$ на поверхности:
\begin{equation}
  n(r) =
  \begin{dcases}
    \frac{2 n_{0}}{ 1 +  \qty(\frac{r}{R})^2}, & r \leqslant R,\\
    n_{0}, & r > R.
  \end{dcases}
\end{equation}
Здесь $R$~--- радиус сферы либо цилиндра. Также
обычно полагают $n_0 = 1$.
Поскольку в методе геометризации на основе квадратичной метрики
диэлектрическая и магнитная проницаемость равны, то запишем:
\begin{equation}
  \begin{gathered}
    \varepsilon^{\crd{i} \crd{j}} = \mu^{\crd{i} \crd{j}},
    \\
    \varepsilon^{\crd{i} \crd{j}} = \varepsilon \delta^{\crd{i} \crd{j}},
    \quad \mu^{\crd{i} \crd{j}} = \mu \delta^{\crd{i} \crd{j}},
    \\
    n = \sqrt{\varepsilon \mu} = \varepsilon = \mu,
    \\
    \varepsilon = \mu = \frac{2}{ 1 +  \qty(\frac{r}{R})^2}, \quad
    r \leqslant R, n_0 := 1.
  \end{gathered}
\end{equation}

Тогда, учитывая~\eqref{eq:tamm:revert:homogeneous:g:tamm:2}, получим следующую метрику:
\begin{equation}
  \label{fig:design:luneburg:g}
  g_{\crd{\alpha} \crd{\beta}} =
  \mqty(\dmat[0]{
    \qty(\frac{2}{ 1 +  \qty(\frac{r}{R})^2})^{-3/2},
    - \qty(\frac{2}{ 1 +  \qty(\frac{r}{R})^2})^{1/2},
    - \qty(\frac{2}{ 1 +  \qty(\frac{r}{R})^2})^{1/2},
    - \qty(\frac{2}{ 1 +  \qty(\frac{r}{R})^2})^{1/2}
  }).
\end{equation}

\begin{figure}
  \centering
  \includegraphics[width=0.4\textwidth]{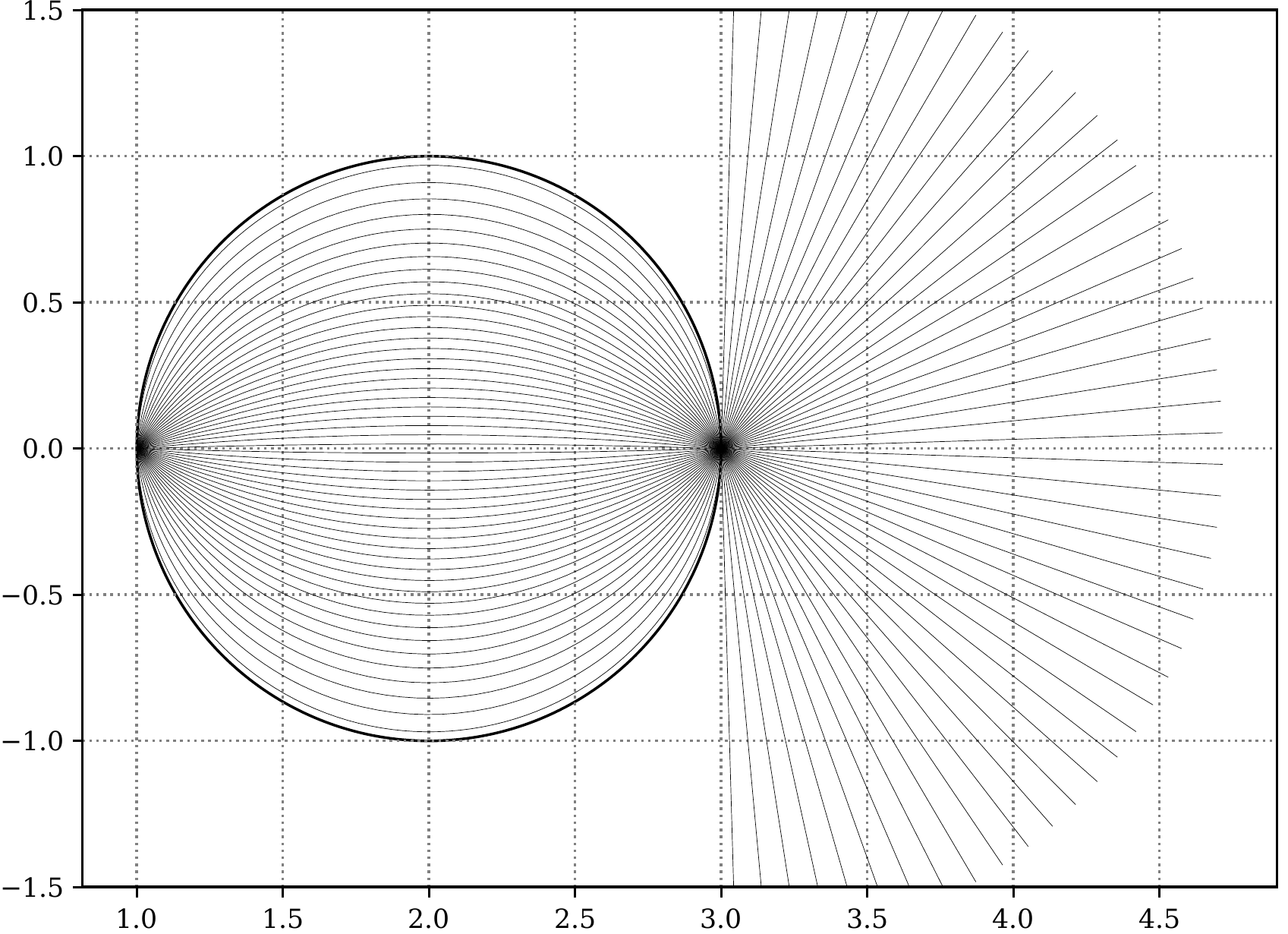}
  \caption{Траектории лучей как геодезические для линзы Максвелла}
\label{fig:design:maxwell:geodesic:rays}
\end{figure}

  Тогда можно изобразить траектории лучей как геодезические в этом
  пространстве (см. рис.~\ref{fig:design:maxwell:geodesic:rays}). Из
  этого рисунка видно, что поведение траекторий лучей совпадает с
  теоретически предсказанным классической
  оптикой~\cite{maxwell:1854:fish-eye}, то есть лучи, выходящие из
  источника на поверхности линзы, собираются в точке, находящейся на
  противоположной поверхности линзы.

  \subsection{Линза Люнеберга}

Линза Люнеберга~\cite{luneburg:1964,morgan:1958:luneberg_lens} есть
градиентная линза. Показатель преломления изменяется в зависимости от
расстояния от центра (сферическая линза) или от оси (цилиндрическая
линза).
При прохождении линзы параллельные лучи фокусируются в одной точке на
поверхности линзы. Лучи, испущенные точечным источником на поверхности
линзы формируют параллельный пучок.

Показатель преломления $n$ меняется от
$\sqrt{2} n_{0}$  в центре до $n_{0}$ на поверхности:
\begin{equation}
    n(r) =
    \begin{dcases}
      n_{0} \sqrt{2-\qty(\frac{r}{R})^2}, & r \leqslant R,\\
      n_{0}, & r > R.
    \end{dcases}
  \end{equation}
  Здесь $R$~--- радиус сферы либо цилиндра. Также
  обычно полагают $n_0 = 1$.

Поскольку в методе геометризации на основе квадратичной метрики
диэлектрическая и магнитная проницаемость равны, то запишем:

\begin{equation}
  \begin{gathered}
    \varepsilon^{\crd{i} \crd{j}} = \mu^{\crd{i} \crd{j}},
    \\
    \varepsilon^{\crd{i} \crd{j}} = \varepsilon \delta^{\crd{i} \crd{j}},
    \quad \mu^{\crd{i} \crd{j}} = \mu \delta^{\crd{i} \crd{j}},
    \\
    n = \sqrt{\varepsilon \mu} = \varepsilon = \mu,
    \\
    \varepsilon = \mu = \sqrt{2-\qty(\frac{r}{R})^2}, \quad
    r \leqslant R, n_0 := 1.
  \end{gathered}
\end{equation}

Тогда, учитывая~\eqref{eq:tamm:revert:homogeneous:g:tamm:2}, получим следующую метрику:

\begin{equation}
  \label{fig:design:luneburg:g}
  g_{\crd{\alpha} \crd{\beta}} =
  \mqty(\dmat[0]{\qty(2-\frac{r}{R})^{-3/4},
    - \qty(2-\frac{r}{R})^{1/4}, - \qty(2-\frac{r}{R})^{1/4}, - \qty(2-\frac{r}{R})^{1/4}}).
\end{equation}

\begin{figure}
  \centering
  \includegraphics[width=0.4\textwidth]{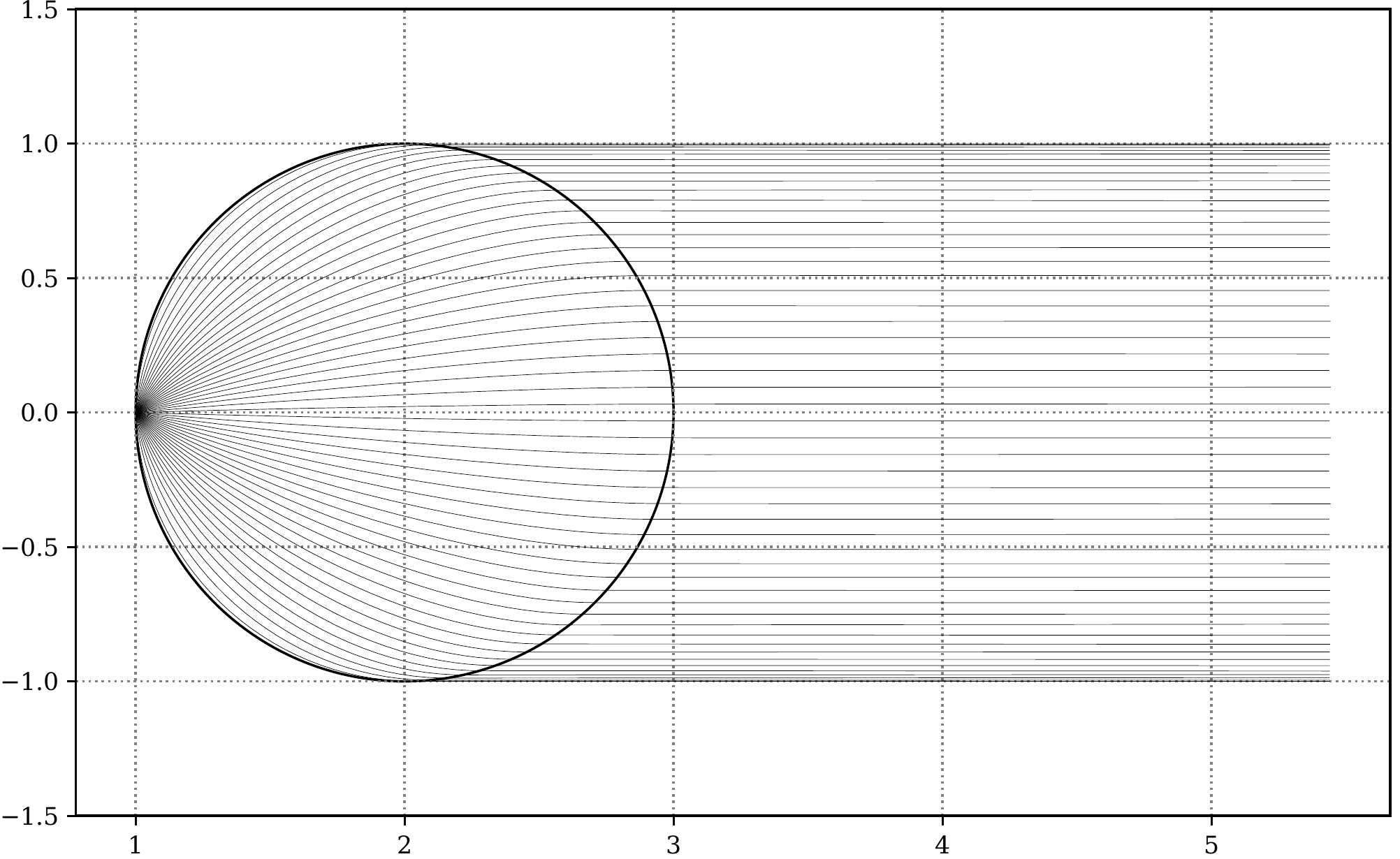}
  \caption{Траектории лучей как геодезические для линзы Люнеберга}
\label{fig:design:luneburg:geodesic:rays}
\end{figure}

Тогда можно изобразить траектории лучей как геодезические в этом
пространстве (см. рис.~\ref{fig:design:luneburg:geodesic:rays}). Из
этого рисунка видно, что поведение траекторий лучей совпадает с
теоретически предсказанным классической
оптикой~\cite{morgan:1958:luneberg_lens}, то есть лучи, выходящие из
точечного источника на поверхности линзы формируют параллельный пучок.

\section{Заключение}
\label{sec:conclusion}

Авторами в работе предложен алгоритм расчёта линз в рамках
геометрического подхода к оптике. Предложены анзацы для случаев
пространственно изотропной и анизотропной сред. На примере широко
известных линз Максвелла и Люнеберга продемонстрировано совпадение
результатов классического и геометрического подходов. Для простоты
расчёты проводились в пределе геометрической оптики.

К сожалению, пока не понятно, даёт ли предлагаемый подход расчёта линз
на основе геометризованной оптики какие-либо преимущества по сравнению
с классической оптикой. Этот
вопрос должен являться предметом дальнейших исследований.

\def\acknowledgmentsname{Благодарности}

\begin{acknowledgments}

Работа частично поддержана грантами РФФИ \textnumero~15-07-08795, 16-07-00556.
Также публикация подготовлена при поддержке программы РУДН <<5-100>>.

\end{acknowledgments}

  \ifthenelse{\boolean{xetex}}
  {\bibliographystyle{ugost2008l}}
  {\bibliographystyle{gost2008l}}

\bibliography{bib/lenses-calculation-algorithm/cite}